\documentclass[sigplan,screen]{acmart} 
 \usepackage{geometry}                		% See geometry.pdf to learn the layout options. There are lots.
\usepackage{graphicx}				% Use pdf, png, jpg, or eps§ with pdflatex; use eps in DVI mode
\usepackage{csquotes}
\MakeOuterQuote{"}
			
\usepackage[frozencache,cachedir=./minted]{minted}
\setminted[haskell]{escapeinside=@@}

\newenvironment{code}
 {\VerbatimEnvironment
  \begin{minted}[fontsize=\small]{haskell}}
 {\end{minted}}

\newenvironment{codex}
 {\VerbatimEnvironment
  \begin{minted}[fontsize=\small]{haskell}}
 {\end{minted}}
 
\newcommand{\hs}{\mintinline[fontsize=\footnotesize]{haskell}}

\newcommand{\hsn}{\mintinline[fontsize=\footnotesize]{text}}

\setcopyright{acmlicensed}
\acmPrice{15.00}
\acmDOI{10.1145/3546189.3549915}
\acmYear{2022}
\copyrightyear{2022}
\acmSubmissionID{icfpws22haskellmain-p11-p}
\acmISBN{978-1-4503-9438-3/22/09}
\acmConference[Haskell '22]{Proceedings of the 15th ACM SIGPLAN International Haskell Symposium}{September 15--16, 2022}{Ljubljana, Slovenia}
\acmBooktitle{Proceedings of the 15th ACM SIGPLAN International Haskell Symposium (Haskell '22), September 15--16, 2022, Ljubljana, Slovenia}

\title[Traversals of Infinite Structures]{A Totally Predictable Outcome: An Investigation of Traversals of Infinite Structures}
\author{Gershom Bazerman}
\affiliation{%
  \institution{Arista Networks}
  \city{San Jose}
  \country{USA}
}


\begin{abstract}
Functors with an instance of the \hs{Traversable} type class can be thought of as data structures which permit a traversal of their elements. This has been made precise by the correspondence between traversable functors and finitary containers (also known as polynomial functors) -- established in the context of total, necessarily terminating, functions. However, the Haskell language is non-strict and permits functions that do not terminate. It has long been observed that traversals can at times in fact operate over infinite lists, for example in distributing the Reader applicative. The result of such a traversal remains an infinite structure, however it nonetheless is productive -- i.e. successive amounts of finite computation yield either termination or successive results. To investigate this phenomenon, we draw on tools from guarded recursion, making use of equational reasoning directly in Haskell. 
\end{abstract}

\begin{document}
\maketitle

\section{Introduction}
Traversable functors, first introduced by McBride and Paterson \cite{mcbride2008applicative}, provide a sort of "complement" to the Applicative type class. While the Applicative type class picks out functors which have a (closed) lax-monoidal structure, the Traversable type class picks out functors which can distribute over this monoidal structure. We recall below the definitions of Applicative and Traversable which we will be generalizing in this paper:

\begin{codex}
class Functor f => Applicative f where
     pure :: a -> f a
    (<*>) :: f (a -> b) -> f a -> f b      
    
class Functor t => Traversable t where
    sequence :: Applicative f => t (f a) -> f (t a)
\end{codex}

Typically the Traversable type class is given by an operation \hs{traverse} rather than \hs{sequence} but the former is given by \hs{traverse f = sequence . fmap f} and the latter is conceptually clearer for our purposes.

By results (independently) of Bird, Gibbons et. al. \cite{bird2013understanding}  and Jaskelioff and O'Connor \cite{jaskelioff2015representation} in a total setting, traversable functors correspond to finitary containers, which is to say that any traversable functor is equivalent to one given as the sum of finite products of its underlying type. However, recursive structures in the Haskell language are not necessarily finitary. For example, the list functor includes lists which have infinite numbers of elements, i.e. streams. Often, as defined in Haskell, traversals naturally operate on such infinite structures. As a trivial example, \hsn{sequence (repeat id) 1}, which makes use of the \hs{Reader} applicative yields an infinite list \hs{[1,1,1...]}.

However, not all applicative functors are amenable to infinite traversals. For example, making use of the \hs{Maybe} monad, \hsn{sequence (repeat (Just 1))} yields $\bot$. Similarly, an infinite sequence making use of the list monad or \hs{ZipList} applicative functor (i.e. sequencing \hsn{repeat [1]} or \hsn{repeat (ZipList [1])} also yields $\bot$.

More interestingly yet, some applicative functors allow traversals that are only productive (non-$\bot$) in some \emph{portion} of their results. For example, making use of the \hs{State} monad:

\begin{codex}
runState (sequence (repeat (modify (+1) >> get))) 0
\end{codex}

yields a tuple whose first projection is the infinite list \hs{[1,2,3...]} and whose second projection is $\bot$.

Algebraic reasoning about functions like these can be tricky, because equational reasoning is usually performed relative to the total, terminating fragment of the Haskell language. Nonetheless, this topic is of particular interest, since as the \hs{State} example shows, infinite traversals very naturally can be used to implement stream transducers -- functions which operate productively over infinite streams, accumulating information and transforming them elementwise, and such transducers are natural ways to present many classic online streaming and incremental algorithms. This paper will draw on the toolkit of guarded recursion to provide a straightforward way of using equational reasoning to consider properties of terms that may be infinite. Then, making use of this kit, we will describe a class of functors, which we term \hs{Predictable} functors, that are suited to performing infinite traversals. Further, we hope the techniques developed in studying this particular question can be made use of more generally in formalizing algebraic reasoning about potentially-non-terminating computations.

The paper is structured as alternating between introduction of new concepts and practical applications of these to the investigation at hand. Section two will introducing a formal \hs{Later} modality in the form of an applicative functor, and demonstrate how this can permit reasoning about total functions over infinite data. Section three will then introduce \hs{Predictable} functors at a high level, which are defined by their relationship to the \hs{Later} modality.  In particular, we can derive the result that for a traversal to be possibly infinite is not a restriction on the structure being traversed, but rather a restriction on the applicative performing the traversal -- in particular such an applicative must commute with \hs{Later} up to a certain notion of equivalence -- loosely speaking this amounts to being free of coproducts.

To reason about extracting properties from guarded recursive reasoning to general code, section four will introduce an inductively defined "evaluation," which is also used to define a form of bisimilarity -- a necessary tool in stating precisely the equational properties of \hs{Predictable} functor. Section five introduces stability, which is a way of thinking about how guarded recursive code can model partiality and potential nontermination. Section six then makes use of this toolkit to investigate predictable functors in more detail, and what operations predictablility is generated by and closed under. Section seven then introduces an extended notion of bisimilarity and evaluation which extends to arrow types, and section eight makes use of this to give a concrete result on the relationship of guarded-recursive infinite traversals and standard traversals in Haskell. 

Finally, section nine applies the theory developed to study many concrete examples of infinite traversals using standard monads, and section ten dwells on some tricky questions that arise from this sort of fine-grained analysis.

\section{Equipping Haskell with a Formal Later modality}

In 2000, Nakhano introduced a typed modal logic for self referential formulae which took semantics in a typed lambda calculus with a modality for guarded recursion \cite{nakano2000modality}. In this logic, types are able to capture when certain data must be evaluated in order for further evaluation to proceed. A term of type \hs{a} which cannot yet be evaluated is given the type \hs{Later a}, expressing that it may be passed around abstractly, but its structure is not yet necessarily available for computation. In this guarded recursive logic, there is necessarily no general function \hs{forall a. Later a -> a}, nor is there a general fixpoint operator. However, guarded recursion is allowed in the form of a function \hs{lfix :: (Later a -> a) -> a}. This says "if you give provide a function that "embeds" a \hs{Later a} into an \hs{a}, then it can be iterated." Such a function allows a fixpoint because the input function, by construction, cannot evaluate the thing it is given, only manipulate it abstractly. Hence, the \hs{Later} modality makes explicit the rule of thumb that programmers who work with the fixpoint operator have already internalized.

In a system like this, necessarily finite and possibly infinite data are distinguished through the use of the \hs{Later} modality. Consider for example, the types
\begin{code}
data List a = 
     LNil 
   | LCons a (List a)
   
data Stream a = 
     Nil
   | Cons a (Later (Stream a))
\end{code}
The former is necessarily finite, as without a general fixpoint operator, no infinite inhabitant can be constructed. However, the latter is possibly infinite. For example, \hsn{lfix (Cons 1)} will produce an infinite stream of ones.

 By an insight of Atkey and McBride \cite{atkey2013productive}, aside from \hs{lfix}, the \hs{Later} modality has exactly the operations of an applicative functor. In particular, we have:
 
 \begin{codex}
 later :: a -> Later a
 lap :: Later (a -> b) -> Later a -> Later b
 \end{codex}
 
Further, in combination these induce functoriality. These operations come with natural intuitions corresponding to the intended meaning of \hs{Later}. "If I know something now, I still know it later. If I will have a function later, and I will have a value later, then later I also will be able to have the result of applying that function to that value."
 
The recognition of \hs{Later} as an applicative functor means that one can program in a fragment of Haskell which enjoys the same properties as a type system with genuine guarded recursion -- i.e. infinite structures can be manipulated in a fashion such that no function yields $\bot$. The recipe for doing so is painfully simple. One simply introduces an abstract datatype equipped with the proper type class instances, and an \hs{lfix} operator, and then does not export its eliminator. 

This is to say one introduces a module supplying the following, and for all code in the fragment we are reasoning about, does not import the constructor \hs{Later}, thus enforcing abstraction.

\begin{code}
newtype Later a = Later a deriving Functor

instance Applicative Later where
  pure = Later
  Later f <*> Later x = Later (f x)

lfix :: (Later a -> a) -> a
lfix f = fix (f . pure)
\end{code}

Further, for all code in sight, one adheres to the discipline of not allowing any direct recursion to occur, and instead using the \hs{lfix} operator uniformly. Working with this sort of discipline without compiler support is not necessarily a good approach for genuine programming in the large. However, for small-scale equational reasoning it suffices, as the condition is straightforward to check by hand.

We call this a "formal" modality because rather than introduce special rules for it into the type-system and semantics of a language, tracking of \hs{Later} is introduced purely formally, as a syntactic device. In the course of this paper, we will refer to terms written with the above discipline (making use of \hs{Later} and \hs{lfix} without explicit recursion) as guarded recursive terms, and generally to code written with this discipline as in the guarded recursive fragment. Additionally, we will refer to datatypes which make use of the \hs{Later} modality (such as \hs{Stream}) as guarded datatypes, types of code written in the guarded recursive fragment as guarded recursive types, and functions written in this fragment (whether or not they are recursive) as guarded recursive functions.

As an example of working in guarded recursive fragment, we present a function for interleaving possibly infinite streams, as well as merging them with a (truncated) zip. We leave it as an exercise to verify that, e.g., the \hs{reverse} function cannot be written.

\begin{code}
sinterleave :: Stream a -> Stream a -> Stream a
sinterleave = lfix $ \f s1 s2 -> case s1 of
  (Cons x xs) -> Cons x (f <*> pure s2 <*> xs)
  _ -> s2

szip :: Stream a -> Stream b -> Stream (a, b)
szip = lfix $ \f s1 s2 -> case (s1, s2) of
  (Cons x xs, Cons y ys) -> 
     Cons (x,y) (f <*> xs <*> ys)
  _ -> Nil
\end{code}

\section{Predictable functors and infinite traversals}

With the above tools in hand, we are ready to provide an introduction to the central idea of this paper. The forward implication of the correspondence between traversable functors and finitary containers is simply the observation that traversable functors are closed under arbitrary sums and finite products \cite{jaskelioff2012investigation}. The goal is to extend traversals to handle guarded types such as \hs{Stream}. Hence, we seek to close extend the class of operations that "infinite-traversable" functors are closed under beyond sum and finite product to also composition with \hs{Later}. This amounts to having (an appropriately lawful) function \hs{Later (t (f a)) -> f (Later (t a))} -- i.e. a function that witnesses that if \hs{t} is infinite-traversable, then so too is the composition of \hs{Later} with \hs{t}. This function in turn decomposes into the composition of \\ \hsn{fmap sequence :: Later (t (f a)) -> Later (f (t a))} with a \\ function \hsn{predict :: Later (f a) -> f (Later a)}. This latter \\ function makes no mention of \hs{t} and so it is effectively not a property of traversable functors, but rather of the applicative functors used to traverse them. This motivates defining a type class, \hs{Predictable} as follows:

% TODO jask has arb sum, finite product. but we have arb sub, sort of arb product? need better grasp on guarded datatypes

\begin{code}
class Predictable f where
  predict :: Later (f a) -> f (Later a)
\end{code}

The inspiration for the name is that it turns out that, loosely speaking, a \hs{Predictable} functor is one where, without inspecting the structure, we can "predict" the outermost constructor. In GHC argot, a predictable functor is one where it is safe to preform an irrefutable pattern match.

In turn, we can now define a candidate type class for infinite-traversable functors:

\begin{code}
class ITraversable t where
  isequence :: (Applicative f, Predictable f) => 
              t (f a) -> f (t a)
\end{code}

Since these classes are for reasoning in the guarded recursive fragment, we require that their instances be given only using tools from that fragment. Further, we require three laws analagous to those for standard traversable functors. Namely:

\begin{itemize}
\item \textbf{Identity}: \\
\hsn{isequence . fmap Identity} = \hs{Identity} \\
     \hs{:: t a -> Identity (t a)}
\item \textbf{Composition}: \\
\hsn{isequence . fmap Compose} = \\ \hs{Compose . fmap isequence . isequence } \\
     \hs{:: t (f ( g a)) -> Compose f g (t a)}
\item \textbf{Naturality}: \\
\hsn{t . isequence} = \hsn{fmap isequence . t} \\
     \hs{:: t (f a) -> g (t a)} 
\end{itemize}

where \hs{t} is a natural transformation between predictable applicative functors, which is to say that it commutes with applicative operations and "weakly" commutes with \hs{predict} (i.e. \hsn{t . predict } is equivalent to \hsn{predict . fmap t} in a sense that will be made precise in the next section).

As an example, here is the \hs{ITraversable} instance for \hs{Stream}.

\begin{code}
instance ITraversable Stream where
  isequence = lfix $ \rec x -> case x of
     Nil -> pure Nil
     Cons a s -> 
         Cons
           <$> a 
           <*> predict (rec <*> s)
\end{code}

Syntactically, it looks nearly the same as the \hs{Traversable} instance for lists, but it makes a judicious use of \hs{predict} to align the \hs{Later} uses in the course of the traversal. As such, straightforward equational reasoning, essentially no different than that for the \hs{Traversable} instance for lists, serves to verify that it satisfies the laws.

In general, for any strictly positive recursive datatype, one can construct a related, "potentially infinite" guarded datatype by syntactically guarding each recursive occurrence by \hs{Later}. Since any \hs{Traversable} datatype can be written as such a strictly positive type, then it follows that for every \hs{Traversable} datatype there is a recursively guarded variant which can be given an instance of \hs{ITraversable}. So while the theory of \hs{Traversable} datatypes, as formulated in sets or the strictly terminating fragment of the lambda-calculus yields a correspondence to finitary containers, those same data types, considered in a richer semantics, need not be finite, and even in that setting where they are not finite, still admit a lawful instance of \hs{ITraversable}.

As a further example of this, we give the a binary tree with labeled leafs, as well as the recursively guarded version of it, and a corresponding \hs{ITraversable} instance for the latter:

\begin{code}
-- "normal" tree
data Tree a = Leaf a | Branch (Tree a) (Tree a)

-- guarded tree
data ITree a = 
     ILeaf a 
   | ITBranch (Later (ITree a)) (Later (ITree a))

-- corresponding infinite traversable instance
instance ITraversable ITree where
  isequence = lfix $ \rec x -> case x of
     ILeaf a -> ILeaf <$> a
     ITBranch x y -> 
             ITBranch 
          <$> predict (rec <*> x)
          <*> predict (rec <*> y)
\end{code}

The next question to consider is when an applicative functor may be given a valid \hs{Predictable} instance. However, this requires specifying what a "valid" predictable instance is. In turn, to do so, some new tools must be developed. The general idea is that our use of types to track guarded recursion in a fragment of Haskell is merely a way of "annotating" existing code to make it make sense. As such, the function \hs{predict} is intended to be merely an accounting device, witnessing some property of the underlying functor, and not to do actual computation. So we might wish to require it be an isomorphism on types. However, in obviously desirable cases it cannot have an inverse. For example, it is easy to send \hs{Later (a -> b)} to \hs{a -> Later b}, but there is no general function going in the other direction. Therefore, we need a notion of equivalence between terms of heterogenous type that suffices to capture the intended semantics -- for this, we introduce something we term bisimilarity by evaluation.

\section{Bisimilarity by evaluation}
We now consider certain equivalence relations on guarded recursive types that are weaker than that given by isomorphism, and which we will need to state more precisely the laws discussed above, and more generally to carry out equational reasoning in the guarded recursive fragment.

As a general motivation, we note that \hs{Later} is not a monad, and in particular, there is no function of the form:
\begin{codex}
forall a. Later (Later a) -> Later a
\end{codex} 

Intuitively, if such a thing existed, it would "collapse" all future timesteps into a single timestep. As such, it would allow unguarded recursion as long as it occurred under at least a single \hs{Later}. 

Nonetheless, we wish to use the guarded recursion modality to reason about terms in Haskell -- a nonstrict language with general recursion. As such, we want to consider equivalence between terms of type \hs{Later a} and \hs{Later (Later a)}, for example. The appropriate notion of equivalence between terms in a nonstrict setting, like that between concurrent terms, should be some form of bisimilarity. In essence, we wish terms to be considered equivalent if under a sequence of abstract observations, each contains the same data, with the same causal ordering dependency. 

The reason we do not want a morphism from \hs{Later (Later a)} to \hs{Later a} is because this would allow a use of \hs{lfix} to perform unbounded recursion. However, for purposes of reasoning about equivalence, rather than calculation, such a map is reasonable -- and indeed, we can go further. For purposes of reasoning up to equivalence, a map \hs{Later a -> a} is reasonable as well. In general since \hs{Later} is a newtype, there will be an associated map from any type with \hs{Later} involved to one without \hs{Later} involved, which, considering both purely as Haskell types, will be an isomorphism. 

The approach we take is that terms are defined in a total setting where there is no general map \hs{Later a -> a} and all recursion is via \hs{lfix}. However, bisimilarity between terms is calculated by "evaluating" these terms to a nonstrict setting with general recursion, and then calculating equivalence using the standard tools of denotational semantics. Rather than constructing a theory of bisimilarity on guarded recursive types intended to capture when they are equivalent in a nonstrict, recursive setting, we simply construct a procedure for evaluating such types to a nonstrict, recursive setting, and directly reason about their equivalence using tools that already exist. This may not seem close to the usual definition of bisimilarity, but it is inspired by the approach of calculating bisimilarity on concurrent processes by reducing the problem to an equivalence relation on their semantics in synchronization trees (the latter relation, in the literature, also often confusingly named bisimilarity).

Continuing the thread of reasoning internally in Haskell, we represent evaluation by means of a type class with an associated type:

\begin{code}
class EvalLater a where
  type Result a 
  leval :: a -> Result a
\end{code}

We require that \hs{Result a} is free of any \hs{Later} in its definition. In keeping with \hs{Later} being a purely formal construction, in turn this notion of evaluation is equally formal -- it does not in fact "evaluate," compute, or beta-reduce terms in any sense. We would also like to require that \hs{leval} is an isomorphism of Haskell types, but for certain instances (such as the \hs{Delay} monad discussed later) this is too strong. Instead, we require that \hs{leval} is an isomorphism of Haskell types only in the case where its domain may not contain an infinite nesting of \hs{Later}, and will revisit the other case in section five. Hence \hs{leval} transports types across a morphism (frequently an isomorphism) that eliminates the use of the \hs{Later} newtype. It may be useful to consider this approach of pairing terms with their semantic meaning via a type class as an internal representation in Haskell of the technique of logical relations.
 
The most important instance is a recursive instance that strips away an outer \hs{Later} and proceeds to continue to evaluate the result. Additionally, all ground types we wish to consider must be equipped with an appropriate instance, and instances for products, sums, and so forth all arise very mechanically. Importantly, this code is not part of the guarded recursive fragment we are reasoning about, but rather belongs to the equational metatheory, and thus freely pattern matches on \hs{Later}. Some example instances are as follows.

\begin{code}
instance EvalLater a => EvalLater (Later a) where
  type Result (Later a) = Result a
  leval (Later x) = leval x

instance EvalLater Int where
  type Result Int = Int
  leval x = x

instance (EvalLater a, EvalLater b) => EvalLater (a, b) 
  where
    type Result (a, b) = (Result a, Result b)
    leval (x, y) = (leval x, leval y)
\end{code}

As  noted early on, when working in the guarded recursive fragment of the language, recursive structures which do not make use of \hs{Later} are already necessarily finite. As such, evaluation on them remains a purely formal operation. For example, for lists we have:

\begin{code}
instance EvalLater a => EvalLater [a] where
  type Result [a] = [Result a]
  leval xs = fmap leval xs
\end{code}

We will refer to structures which do not make use of \hs{Later} either directly or indirectly as \emph{finite} with regards to the guarded recursive fragment.

The only slight complication arises in the case of guarded datatypes, which may themselves make use of \hs{Later}. In such a case, the need for an associated type becomes very clear, as we are evaluating to an entirely different type, which we have established is isomorphic, rather than to a type obviously syntactically related to the given type. For example, partially infinite streams may be evaluated like so:

\begin{code}
instance EvalLater a => EvalLater (Stream a) where
  type Result (Stream a) = [Result a]
  leval Nil = []
  leval (Cons x  xs) = leval x : leval xs
\end{code}

In general, this construction allows that every recursively guarded variant of a strictly positive datatype can evaluate to the original type.

Functors all enjoy the property that \hs{leval = lt . fmap leval}, where \hs{lt} is an appropriate natural transformation performing "outer evaluation" -- i.e., evaluation on functors factorizes into a parametric evaluation on the "shape" of the functor and the \hsn{fmap} of an elementwise evaluation on the contents.  As such while \hs{Result} is only defined on things of kind \hs{*} we also will refer to \hs{Result f} where \hs{f} is a functor. Finite functors enjoy the further property that \hs{lt} is \hs{id} -- which is to say that the outer shape of a finite functor is already free of \hs{Later}, and so evaluation of such a functor need only evaluate its contents.

We may now define two terms \hs{x} and {y} of types {a} and {b} as \emph{evaluated-bisimilar}  when there is an isomorphism \\ \hs{f :: Result a -> Result b} and further \hsn{f (leval x)} = \hs{leval y}. For our purposes, this isomorphism of evaluated terms will typically be \hs{id}. 

There is thus a sense in which guarded recursive types may be seen as fibered over isomorphism classes of standard Haskell types, providing for each such class a (partially ordered) set of possible refinements of evaluation dependency. This is to say that many guarded recursive types may evaluate to the same (isomorphism class of a) type, but those guarded recursive types will themselves not be necessarily isomorphic as guarded recursive types -- instead, they will be partially ordered by how many inhabitants of that evaluated type they represent. For example, the types \hs{Stream a}, \hs{Later [a]}, and \hs{Later (Later [a])} all "live above" \hs{[a]}, but more lists may be represented by \hs{Stream a} than by \hs{Later [a]}.

It is worth noting that evaluation is defined on all terms, not just those in the guarded-recursive fragment, and consequently evaluated-bisimilarity as a relation can extend not only to guarded recursive terms, but to general terms.

When \hs{leval} is an isomorphism, the isomorphism of an evaluated-bisimilarity extends to a productive isomorphism between the two bisimilar terms themselves, and not just their \hs{Result}s. However this isomorphism need not be itself a guarded recursive function. Because we wish to state identities in the guarded recursive fragment relative to functions that themselves are guarded recursive, and to cover cases where \hs{leval} is not necessarily an isomorphism, we introduce a further notion.

A guarded recursive function \hs{a -> b} is a guarded weak bi-equivalence (\emph{gwbeq}) if it sends a term to an evaluated-bisimilar term. The name is justified because all identities are clearly gwbeq and further, it is straightforward to show that the class of gwbeqs is closed under composition and furthermore satisfies "two out of three" -- that is if  \hs{f :: a -> b} and \hs{g :: b -> c} and two out of three of $\{f, g, g \odot f\}$ are gwbeq, then the third is as well. It is important to note that a gwbeq of type \hs{a -> b} will typically not have a corresponding inverse gwbeq of type \hs{b -> a} -- though in the next section, we will consider the special case where this does occur.

We will also at times consider bi-equivalences which are not necessarily written in the guarded recursive fragment, and will refer to them as evaluated bi-equaivalences (\emph{ebeq}). By definition, \hs{leval} itself is an ebeq.
 
We can now state the desired law of a predicable functor -- for a valid instance of \hs{Predictable}, the guarded recursive function \hs{predict :: Later (f a) -> f (Later a)} must be a guarded weak bi-equivalence.
  
As an example of using gwbeqs in reasoning, we now prove a further lemma. By construction and definition, a \hs{Predictable} functor can commute with all \hs{ITraversable} functors. Conversely, we can show that if a functor \hs{f} can commute with all \hs{ITraversable} functors, then that functor is necessarily predictable. In particular, \hs{Later} is itself straightforwardly \hs{ITraversable}. Instantiating \hs{isequence} at \hs{Later} yields \hs{Later (f a) -> f (Later a)} -- precisely the type of \hs{predict}. By identity and naturality, we can conclude that this must be a gwbeq, which shows the function has the desired properties. Thus we see that being \hs{Predictable} is both necessary and sufficient for an \hs{Applicative} functor to yield infinite traversals.
 
\section{Productivity and Stable values}
The goal of this section is to develop a notion of only possibly productive types in guarded recursion, which will be made use of in cataloguing instances of \hs{Predictable} functors. It is tempting to describe working in a guarded recursive system as a situation "where the types ensure everything you write is productive" -- but this is a misleading intuition. The types ensure that no pattern match on any term will yield $\bot$. However, it is possible to encode data that in the "evaluated" semantics absolutely does yield $\bot$ -- for example, an infinite nesting of \hs{Later}. The point is that, in doing so, the lack of productivity can be read off directly from the types. So guarded recursion is a situation where reasoning about productivity is made manifest at the type level.

More generally, it is useful to to work with types which are only possibly productive -- i.e. where there may be arbitrary sequences of \hs{Later} applications. These can be captured by a \hs{Delay} type, following Capretta \cite{capretta2005general}.

\begin{code}
data Delay a = Now a | Wait (Later (Delay a))
\end{code}

Using this type, one can, for example write a function to compute the last element (should it exist) of a possibly infinite stream.

\begin{code}
slast :: Stream a -> Delay (Maybe a)
slast = go Nothing
  where go = lfix $ \f def s1 ->
         case s1 of
          (Cons x xs) -> Wait $ f <*> pure (Just x) <*> xs
          Nil -> Now def
\end{code}

The same idea can also be used to capture other structures that are possibly productive, such as possibly productive infinite streams:

\begin{code}
data PStream a =
      PNil
    | PWait (Later (PStream a))
    | PCons a (PStream a)
\end{code}

Corresponding \hs{EvalLater} instances can also be written for possibly productive structures, such as below:

\begin{code}
instance EvalLater a => EvalLater (Delay a) where
  type Result (Delay a) = Result a
  leval (Now x) = leval x
  leval (Wait x) = leval x

instance EvalLater a => EvalLater (PStream a) where
   type Result (PStream a) = [Result a]
   leval PNil = []
   leval (PWait x) = leval x
   leval (PCons x xs) = leval x : leval xs
\end{code}

The \hs{leval} in these instances is clearly not an isomorphism of Haskell types. Further, there is no particular rule that can uniquely determine what the implementation should be. For example, something isomorphic to \hs{Delay ()} could also be intended to represent a lazy Peano numeral -- and that would have a significantly different \hs{leval} implementation. The most that can be required is that the given function have a right inverse. Beyond that, the range of possible implementations of \hs{leval} is handled with regards to equational properties by imposing a constraint of "bisimulation-invariance" regarding the functions we reason over. Both the right inverse and the invariance constraint will be treated further in section seven.

As discussed earlier, a gwbeq, unlike an isomorphism, may not be invertible -- i.e. there is always a map \hs{a -> Later a} and there is no general guarded recursive map in the other direction, either as an actual retract, or even as a gwbeq. Possibly productive structures correspond to cases where such a gwbeq \emph{does} exist. Inspired by the terminology of "stable propositions" in modal logic we term types that permit such a map "Later-stable" or just "stable". These will play a special role in the construction of certain classes of predictable functors, and can be equipped with a type class as follows:

\begin{code}
class Stable a where
   wait :: Later a -> a

instance Stable (Delay a) where
   wait = Wait

instance Stable (PStream a) where
   wait = PWait
\end{code}

% As a convention, we will term types which nowhere in them may contain infinite nestings of \hs{Later} as "productive types". A type such as \hs{(Int, Delay Int)} is neither stable nor productive -- it is not stable because it cannot contain \emph{solely} an infinite nesting of \hs{Later}, but it is also not productive, because it may contain such a nesting at some place within it.

As an aside, a gwbeq sending a term to a term that is of a stable type may be termed a "stabilization" and consequently, for any type \hs{a}, \hs{Now :: a -> Delay a} is the universal (terminal) stabilization. 

\section{Examples of predictable functors}
Here, we consider some examples of predictable functors, and identify a large class of valid instances. As the instances below show, \hs{Predicable} is closed under composition, and product. Base instances that generate it include exponentiation by a constant (i.e. a function from a type not containing \hs{a}), and \hs{Later} itself.
\begin{code}
instance Predictable Later where
  predict = id

instance Predictable ((->) r) where
  predict x = \y -> fmap ($ y) x
   
instance (Predictable f, Predictable g, Functor f) => 
               Predictable (Compose f g) where
  predict = Compose . fmap predict 
               . predict . fmap getCompose

data Prod f g a = Prod {pr1 :: f a, pr2 :: g a}

instance (Predictable f, Predictable g) => 
               Predictable (Prod f g) where
  predict x = Prod 
                  (predict (fmap pr1 x)) 
                  (predict (fmap pr2 x))
\end{code}

Strikingly, \hs{Later} is not closed under sum. In particular, consider the type:

\begin{codex}
data Choice a = C1 a | C2 a
\end{codex}

The outermost constructor of the result of \hs{predict} would necessarily be \hs{C1} or \hs{C2} -- however, by the construction of the \hs{Later} modality, there is no mechanism in the guarded recursive fragment to determine which one. In other words, no progress can be made because the choice of outermost constructor is not \emph{predictable}. 

Nonetheless, this restriction does not mean that \hs{Predictable} structures are not closed under product with any constant -- in fact, they remain closed precisely under multiplication by \emph{specific} constants -- those which are stable. In particular:

\begin{code}
instance (Stable c) => Predictable ((,) c) where
   predict :: forall a. Later (c, a) -> (c, Later a)
   predict x = (wait (fst <$> x), snd <$> x)
\end{code}

As the type signature for \hs{predict} indicates, the function needs to move the type \hs{c} out of the later modality -- which is exactly what is permitted by it being \hs{Stable}.

The above instance in fact extends to a bi-implication --
 \hs{((,) c)} is predictable if and only if \hs{c} is stable. In particular, if \hs{Later (c, a)} has a gwbeq to \hs{(c, Later a)} then so too does \hs{(Later c, Later a)}, and this means that \hs{predict} implies a morphism \hs{Later c -> c} that is a gwbeq. The last is, by definition, stability.

As a loose intuition, if traversable functors are presentable as polynomials, then strictly-positive predictable functors are those which are presentable as polynomials concentrated in a single degree -- i.e. of the form $s*X^c$ with s a stable constant, and c an arbitrary constant.

However, we note that it is possible (though as we will see later, not necessarily useful) to give a predictable instance to a not-strictly-positive datatype as well. In particular, we have, for any contravariant functor \hs{f}:

\begin{code}
predictNegative :: (Contravariant f, Stable r) => 
     Later (f a -> r) -> f (Later a) -> r
predictNegative f = 
     \z -> wait $ fmap ($ contramap pure z) f
\end{code}

Both the contravariant instances and the multiplication instances provide examples of \hs{Predictable} functors that are not \hs{Representable} (i.e. of the form $X^c$). Further, being \hs{Predictable} does not imply being \hs{Applicative}. For example, $(s,a)$ is only applicative with a specified choice of monoidal action for \hs{s}. A stronger counterexample generated by this would be picking an \hs{s} which is uninhabited (and thus trivially stable).

\section{Bisimilarity by evaluation for arrow types}

The constructions thus far describe properties holding in the guarded recursive fragment. However, an important goal is to  be able to extract reasoning from this fragment and apply it to general terms, and in particular to establish not only that \hs{isequence} can be defined productively, but the conditions under which a \hs{sequence} is productive. To do so we need to extend further our toolkit for reasoning, and in particular, to extend the notion of evaluated bisimilarity to cover types which contain arrows. When we wish to consider evaluation on the function space, the need arises to go in the other direction as well -- that is, to \emph{lift} general values (which are posited to satisfy appropriate productivity conditions) back into the guarded recursive realm. We therefore introduce the following type class to represent inductively the desired inverse: 

\begin{code}
class EvalLater a => LiftLater a where
  llift :: Result a -> a
\end{code}

We require that \hs{llift} be an ebeq, and in particular, a right-inverse of \hs{leval}. This is to say that \hs{leval . llift} should be the identity.

By recursive definition, the \hs{EvalLater} class allows for a meaningful notion of evaluation on arrow types as follows:

\begin{code}  
instance (LiftLater a, EvalLater b) => 
  EvalLater (a -> b) where
     type Result (a -> b) = Result a -> Result b
     leval f = leval . f . llift

instance (LiftLater a, LiftLater b) => 
   LiftLater (a -> b) where
     llift f = llift . f . leval
\end{code}

Inverses (up to bisimilarity) for various defined evaluations follow mechanically. As some examples:

\begin{code}
instance (LiftLater a) => LiftLater (Later a) where
  llift = Later . llift

instance (LiftLater a) => LiftLater (Delay a) where
  llift = Now . llift

instance (LiftLater a, LiftLater b) => 
   LiftLater (a, b) where
     llift (x, y) = (llift x, llift y)

instance LiftLater Int where
  llift = id
  
instance LiftLater a => LiftLater (Stream a) where
  llift [] = Nil
  llift (x:xs) = Cons (llift x) . Later $ llift xs
\end{code}

It is important to note that these inverses do not, and cannot, send general Haskell terms to guarded recursive terms which are guaranteed productive. In particular, a guarded recursive construction of infinitely nested \hs{Delay} is perfectly reasonable, and under evaluation it goes to $\bot$. Lifting this back in turn does not land in guarded recursive terms (since such terms do not directly contain $\bot$) but rather in terms which make use of \hs{Later} but exist in general Haskell semantics -- so the lifting only goes "halfway" back. However, this halfway back intermediate form is all that is necessary for the bookkeeping of the ultimate goal, which is simply that \hs{leval} be defined properly on function types.

In the following sections, we will assume that all types in sight have been equipped with appropriate \hs{LiftLater} and \hs{EvalLater} instances. Further, we note that computing with \hs{LiftLater} is reasonable, but due to injectivity constraints in GHC's checker, \hs{EvalLater} is a class for reasoning with, but much less so computing with.

It is important to note that the ability to evaluate functions does not mean that all algebraic properties of a guarded recursive function necessarily descend to its evaluation. Loosely speaking, \hs{leval} is not necessarily functorial -- that is to say \hsn{leval f (leval x)} need not be equal to \hsn{leval (f x)}. For example, consider a function of type \hs{Delay Int -> Bool} that returns \hs{True} if the outermost constructor is \hs{Now} and \hs{False} otherwise. The action of this function under \hs{leval} would, on the other hand, always yield \hs{True}. The problem is that the function can detect "how evaluated" its input is, while its evaluation, by construction cannot. The class of functions that do commute with \hs{leval} are those which do not distinguish between "layers" of \hs{Later} -- i.e. which take bisimilar inputs to bisimilar outputs. We call such functions bisimulation-invariant, and  bisimulation-invariant properties of bisimulation-invariant functions descend to their evaluations, as one would desire.

% TODO leval of a function that is not bisimilar-invariant is still definitionally an isomorphism, but somehow does not respect it -- specify how

\section{Predictable functors yield productive traversals}

% TODO define "contains bottom" on guarded recursive fragment, 

% productivity in Later (causal) does permit bottom in a sense -- but it is a step indexed bottom.

% Stable elements do not evaluate to necessarily not-containing bottom things.

% Full theorem holds for finite predictable applicatives -- products (which are also exponentiation by a constant) and Nothing else
% More limited for predictable productive applicatives -- productive is there is always an algebra (f a -> a) for all a. Delay is not productive.
% Cont is _not_ productive
% Actually holds if we require an algebra for all a on the guarded recursive fragment.
% Maybe just require strict positivity, show weird example.
% cont is not productive but cont over writer is productive?

% Need to be more precise with what "productive" means -- means doesn't get stuck, doesn't mean you can't define thigns that don't return a usable value. Just... the evaluation is step indexed. "causal"

The goal of this section is to explicitly establish that the productive properties of infinite-traversals in the guarded recursive fragment yield related properties which are generally useful for traversals even when considering only "normal" Haskell code (without a \hs{Later} in sight). In particular, we will show that all \hs{Traversable} functors admit \hs{Traversable} instances which are productive on predictable applicatives.

We set out to establish when one can "infinite-traverse" a functor productively. Let us now define productive traversal more precisely. Productiveness is typically only defined on functions yielding coinductive structures, and is the property that any finite sequence of operations on the result of such a function will terminate in finite time. However, for a traversal \hsn{t (f a) -> f (t a)}, the applicative functor \hs{f} that is the action of the traversal is not necessarily a coinductive structure. Further, it may yield something with a productive \hs{t a} but coupled with a non-terminating component. So some subtlety is required in defining productivity. For the purposes of this paper, we say that a traversal \hs{h} of type \hsn{t (f a) -> f (t a)} is productive if \hs{f} has a universal algebra \hs{alg :: forall a. f a -> a}, and for an \hs{x} that does not contain $\bot$, no finite sequence of pattern matches on \hsn{alg (h x)} yields $\bot$. For example, if we traverse a stream with a \hs{Writer} monad, the result is a pair of the accumulated writes, and the stream. The traversal is productive, because the algebra yields the second projection (the stream), which is productive in the typical sense -- and this holds true, even though the first projection (the result of a potentially infinite sequence of monoidal writes) may itself be $\bot$.

Further, we call a guarded traversal productive if the algebra is itself in the guarded fragment, and further, for an \hs{x} that is productive (can contain no infinite sequence of \hs{Later}), then \hsn{alg (h x)} is likewise productive. By the properties of guarded recursion, we can "read off" the productivity of a guarded traversal simply from checking that \hs{h} is itself productive. It is straightforward to check that if a bisimulation-invariant guarded traversal \hs{h} is productive in the guarded recursive sense, then the evaluated traversal \hs{leval h} is productive in the sense we introduced above.

Earlier, we sketched how any \hs{Traversable} functor generates a recursively guarded variant which is an \hs{ITraversable}. Here, we consider the converse. Given any \hs{ITraversable} functor \hs{t}, then for any predictable applicative functor \hs{f}, there is a function \hs{seq :: Result t (f a) -> f (Result t a)}, given by \hsn{leval isequence}. Because \hs{leval} is an ebeq, then one can check that the \hs{ITraversable} laws descend to the appropriate \hs{Traversable} laws. At a high level, everything in sight is either removing or adding \hs{Later}, and since \hs{Later} is invisible under \hs{leval} and all functions we consider are bisimulation-invariant, then properties established in the guarded recursive fragment hold under evaluation. Notably, among these properties is productiveness itself. Thus, we can conclude that all \hs{Traversable} functors admit \hs{Traversable} instances which are productive on predictable applicatives.

As a fully worked example, we consider traversals of streams by finitary functors, reasoning equationally, first fusing the evaluation of the result of the sequencing, and then fusing the input to the sequencing.

The initial step makes use of the fact that \hs{f} is a finitary functor. We then substitute the definition of \hs{lfix}. Then, we push the application of \hs{leval} through the case statement, and make use of the fact that \hs{predict},  \hs{llift} and \hs{leval} are all ebeq to substitute between them. Finally, we compose with \hs{llift} and calculate -- which amounts to substitution of pattern matching, changing the type of the recursive call, and elimination of the inner \hs{leval}.

\begin{codex}
sequenceS :: (Applicative f, Predict f) =>
           Stream (f a) -> f (Stream a)
sequenceS = lfix $ \rec x -> case x of
     Nil -> pure Nil
     Cons a s -> 
         Cons 
           <$> a 
           <*> predict (rec <*> s)
           
leval . sequenceS :: Stream (f a) -> f [a]
\end{codex}
$=$ substitution of \hs{sequenceS} and \hs{lfix}
\begin{codex}
fmap leval . fix $ \rec x -> case x of
     Nil -> pure Nil
     Cons a s -> 
         Cons  
           <$> a 
           <*> predict (pure rec <*> s)
\end{codex}
$=$ \hs{leval} through \hs{fix}, inserting matching \hs{llift}.
\begin{codex}
fix $ \rec x -> fmap leval $ case x of
     Nil -> pure Nil
     Cons a s -> 
         Cons  
           <$> a 
           <*> predict (fmap llift . rec <$> s)
\end{codex}
$=$ \hs{leval} through case, rewriting \hs{predict} and \hs{fmap llift}
\begin{codex}
fix $ \rec x -> case x of
     Nil -> pure []
     Cons a s -> 
         (:) 
           <$> a 
           <*> leval (rec <$> s)
\end{codex}           
Finally, we compose with \hs{llift} and calculate:
\begin{codex}           
leval . sequenceS . llift :: f [a] -> f [a]
\end{codex}
$=$ substitution and reduction
\begin{codex}
fix $ \rec x -> case x of
     [] -> pure []
     (:) a s -> 
         (:) 
           <$> a 
           <*> rec s
\end{codex}

This last term is the standard \hs{sequence} for lists, and hence we can conclude that the \hs{Traversable} instance for lists is productive on predictable applicatives (when they have the corresponding algebra necessary to state the property).

% Delay and Guard are an adjoint monad/comonad pair
% make Guard not strong using the Static trick

\section{Examples (and nonexamples) of predictable functors and their traversals}

\begin{comment}
\begin{code}

class Contravariant f where
  contramap :: (a -> b) -> f b -> f a

instance Semigroup w => Semigroup (Delay w) where
  Now x <> Now y = Now (x <> y)
  x <> Wait y = Wait $ fmap (x <>) y
  Wait x <> y = Wait (fmap (<> y) x)

instance Monoid w => Monoid (Delay w) where
  mempty = Now mempty
  
\end{code}
\end{comment}

We now revisit the examples from the introduction, showing how the machinery developed can aid in reasoning about infinite traversals. As a general theme, there are multiple guarded recursive types which lie over every standard type, which may be viewed as "causal refinements" of the underlying type. While all predictable types yield productive traversals, not all capture equally granularly the full properties of the underlying applicative which they are modeling, and there are examples which show that there is not necessarily a "single best" refinement for all uses.

As a warm up, the \hs{Reader} monad is straightforwardly a predictable functor, and so is productive on infinite traversals.

\begin{code}
instance Predictable (Reader r) where
  predict x = reader $ \r -> fmap (($ r) . runReader) x
\end{code}

The \hs{Writer w} monad introduces a complication -- to specify a \hs{Predict} instance, the monoid carried by the writer must be stable:.

\begin{code}
instance (Stable w, Monoid w) => 
         Predictable (Writer w) where
  predict x = writer $ 
      (fst . runWriter <$> x, 
       wait $ snd . runWriter <$> x)
\end{code}

As an example, when \hs{w} is a monoid, \hs{Delay w} has a natural monoid structure and so we have the fact that in general, \hsn{runWriter . sequence} (of type \hs{[Writer w a] -> ([a], w)}) will yield a non-$\bot$ second projection only on the occasion that the input list is finite.

However, for specific choices of \hs{w} there are more precise types possible. For instance, there is a straightforward Haskell instance of monoid for \hs{PStream a} -- the type of possibly productive streams.

\begin{code}
instance Semigroup (PStream a) where
   PNil <> x = x
   PCons x xs <> y = PCons x (xs <> y)
   PWait xs <> y = PWait ((<>y) <$> xs)
  
instance Monoid (PStream a) where
   mempty = PNil
\end{code}

The resultant \hs{Writer (PStream a)} expresses that when the accumulator of a writer is itself a list, then sequencing an infinite list can yield a "semi-productive" second component -- in particular, if there are $m$ writes within the first $n$ terms of the list, then sequencing up to the first $n$ terms of the list will yield $m$ results in the second component.

The \hs{State} monad presents difficulties similar to those of \hs{Writer}, but slightly more complicated. The most naive choice for a \hs{Predict} instance would be \hs{State (Delay s)}. But then at any individual point in a computation, the "real" state would be guarded by a \hs{Delay} operator and so appear not necessarily productively accessible. The property we want to capture is more subtle -- on an infinite list, the second projection of \hs{runState} is indeed $\bot$, as there is no "final" state. However, at any individual point in a \hs{State} computation, the state \emph{thus far} is immediately accessible. A more granular type to capture the productivity of \hs{State} as well as its algebraic operations can be achieved by moving to Ahman and Uustalu's and algebraic generalization of state into the pair of a reader and writer monad -- which they term the \hs{Update} monad \cite{ahman2014update}. The update monad decouples the information on the left and right hand side of the function arrow -- one still reads the state, but writes are not directly of the state, but instead of a monoidal output which \emph{acts} on the state. This decoupling solves the difficulties in presenting a state monad that has good \hs{Predictable} properties -- only the output need be stable, and not the state itself. This does not yield the \hs{State} monad per-se, but does provide a reasonable simulation therein.

We recall briefly the \hs{Update} monad below, and present the appropriate \hs{Predictable} instance.

\begin{code}
data Update p s a = Update {runUpdate :: s -> (p, a)}
    deriving Functor
    
class (Monoid p) => ApplyAction p s where
   applyAction :: p -> s -> s

instance (ApplyAction p s) => Monad (Update p s) where
  Update u >>= f =
    Update $ \s ->
      let (p, a) = u s
          Update t = f a
          (p', a') = t (applyAction p s)
       in (p <> p', a')

putAction :: p -> Update p s ()
putAction p = Update $ \_ -> (p, ())

getState :: Monoid p => Update p s s
getState = Update $ \s -> (mempty, s)

instance Stable p => Predictable (Update p s) where
  predict x = Update $ 
      \s -> predict (($ s) . runUpdate <$> x)
\end{code}

With this in hand, we can give apply actions to appropriate stable structures. For example, the "state-logging" monad of Piróg and Gibbons \cite{pirog2013monads} can be instantiated as the action of the heads of partially productive streams.

\begin{code}  
instance ApplyAction (PStream a) a where
  applyAction (PCons x _) _ = x
  applyAction _ s = s
\end{code}  

By using "head" instead of "last", unlike the version in Ahman and Uustalu (example 5 of \cite{ahman2014update}), this allows infinite traces. The result of an infinite traverse is a pair, whose first component is the semi-productive stream of all intermediate states, and whose second component is the necessarily productive stream of every traversed result. There is an important sublety to this argument -- the instance given for \hs{PStream} is not bisimulation invariant! However, it is invariant in the lucky circumstance that the only actions which are applied are those whose first element is not delayed. Fortunately, the infinite traversals of \hs{Stream} and \hs{ITree} we have considered thus far obey this property, which we term being \emph{prompt}, and will consider further in the next section. At this point, we consider it an open question whether a guarded-recursive and predictable version of \hs{State} is possible such that all its operations can be written in a bisimilar-invariant fashion, though, from the discussion in the next section, it seems unlikely.

We now consider a few nonexamples. As one would hope, because they are sum types, applicative functors such as \hs{Maybe} and \hs{Either a} cannot be given a \hs{Predictable} instance, which corresponds to the fact that every element of an infinite sequence would need to be inspected before the head of a traversal could be produced -- i.e. a \hs{Nothing} anywhere in an infinite stream would render the entire traversal to be \hs{Nothing}. 

A similar situation pertains with lists, which are also constructed as sum types, and so not \hs{Predictable}. In this case it corresponds to the fact that, for the list (nondeterminism) monad, a sequence amounts to a cross product -- to calculate if the first element of the first list in the result of a sequencing is nil requires determining if any element anywhere in the sequenced list is nil. If sequencing is instead performed with the ZipList applicative, we arrive at a truncated transposition, but the same constraint applies. The same constraint also applies to the lifting of lists to possibly infinite streams. However, necessarily infinite streams, which are isomorphic to functions from the naturals, do gain the natural \hs{Reader} predictable structure, and further can be observed to not be constructed using sum types. As such, an infinite sequencing can be written to transpose an infinite stream of infinite streams, which amounts to a transposition of an infinite array. Necessarily infinite streams only admit a "zip"-like applicative and not a cross-product monad (the unit for the latter is necessarily a finite list). As an unfortunate, but understandable, consequence, the techniques developed here do not permit the generation of all possible strings of bits by something such as \hs{sequence (repeat [True, False])}.

Finally, we consider the slightly subtle problem of the continuation monad. The continuation monad in fact admits a predictable instance, as given below:

\begin{code}
data Cont r a = Cont {runCont :: (a -> r) -> r}

instance Stable r => Predictable (Cont r) where
  predict x = Cont $ \z -> 
      wait $ fmap ($ (z . pure)) (runCont <$> x) 
\end{code}

However, there is no general algebra \hs{alg :: Cont r a -> a}, and further, even when \hs{r} and \hs{a} are the same type, the stability requirement on \hs{r} and the action of \hs{predict} (which always inserts a wait over the entire structure) means that the result of an infinite traversal is necessarily infinitely delayed. So this gives a situation where an infinite traversal is well-defined in the guarded-recursive fragment, but this traversal is nevertheless not productive in the guarded-recursive sense or in the sense of evaluated semantics.

\begin{comment}
\begin{code}
  
foo :: [(Update (PStream Int) Int Int)  ]
foo = [getState, putIt 1 >> getState, putIt 3 >> getState, getState >>= \x -> putIt (x + 1) >> getState, getState, getState >>= \x -> putIt (x + 1) >> getState, getState >>= \x -> putIt (x + 1)  >> getState >>= \x -> putIt (x + 1)  >> getState >>= \x -> putIt (x + 1)  >> getState]
   where putIt i = putAction $ PCons i PNil
  
bar :: [State Int Int]
bar = [get, put 1 >> get, put 3 >> get, get >>= \x -> put (x + 1) >> get, get, get >>= \x -> put (x + 1) >> get, get >>= \x -> put (x + 1) >> get >>= \x -> put (x + 2) >> get]

listToStream (x:xs) = Cons x . Later $ listToStream xs
listToStream [] = Nil

streamToList Nil = []
streamToList (Cons a (Later b)) = a : streamToList b

deriving instance Show a => Show (PStream a)

deriving instance Show a => Show (Stream a)

deriving instance Show a => Show (Later a)

instance (ApplyAction p s) => Applicative (Update p s) where
  pure a = Update $ \_ -> (mempty, a)
  x <*> y = x `ap` y
\end{code}
\end{comment}

\section{Prompt traversals, sequencing, and bi-infinite structures}

A consequence of the analysis of traversals in Bird, Gibbons, et al. \cite{bird2013understanding} is that traversals of a datatype correspond to permutations of its elements and in fact there are as many traversals of a given datatype as there are permutations on its arities. There is a sense in which this holds true on infinite traversals, but also a sense in which it breaks down. In particular, it holds somewhat true only as long as all functions in sight are bisimulation-invariant. However, as soon as any function (such as the action of the update "simulation" of \hs{State} above) breaks bisimulation-invariance (through "detecting" a \hs{Later}), then the traversal of a structure making use of such functions no longer necessarily corresponds to a traversal in the evaluated semantics. At first, this may seem like a rather dreary state of affairs -- but it has good cause.

Consider, for example, the "backwards" sequence on a list given as:

\begin{code}
backquence :: Applicative f => [f a] -> f [a]
backquence (x:xs) = flip (:) <$> backquence xs <*> x
backquence [] = pure []
\end{code}

This is a valid traversal, and is infinite productive on, for example, \hs{Reader} and \hs{Writer}. On \hs{State} it will yield $\bot$ on an infinite list. This corresponds to the fact that \hs{State} with a \hs{get} that actually can access the (undelayed) state, is not in fact predictable. The predicatable approximation of \hs{State} that via \hs{Update} simulates "get" through a function that is not bisimulation-invariant except in the circumstance that a traversal is prompt -- precisely what is violated here.

A prompt traversal is defined, syntactically, as one in which every \hs{predict} is associated to the far right of an applicative chain. Consider the guarded-recursive version of backquence, as given below:

\begin{code}
ibackquence :: (Applicative f, Predictable f) 
                     => Stream (f a) -> f (Stream a)
ibackquence = lfix $ \rec x -> case x of
  Nil -> pure Nil
  Cons a s -> flip Cons <$> predict (rec <*> s) <*> a
\end{code}

The \hs{predict} call occurs in a term to the left of an \hs{<*>} where the right hand does not contain a predict call. So this traversal is not prompt. Conceptually, we have secretly been working with a form of "Later-bias" in our monoidal functors. Predictable monoidal functors that contain a product with a stable type use that stable type to "contain" \hs{Later} within them. The monoidal nature of applicative functors typically descends to a monoidal action on that type. The monoids we have considered thus far have been, as is typical, "left-strict-biased" -- this is to say that given \hs{(x <> y)}, the first non-later information will be nested in as many laters as induced by the \emph{left}-hand side of the append. Because predict calls force another layer of later-nesting, placing them on the left side of applicative application "inverts causality," and in particular disrupts the ability of non-bisimulation-invariant functions to access an undelayed first piece of information from our carrier monoid.

The "bias" of monoids we discuss here is not purely an operational quirk. It corresponds to an important mathematical fact -- a monoidal operation is finitely associative, but not necessarily infinitely so. In particular, we can consider the bias-preserving liftings of the \hs{Last} and \hs{First} monoids to stable structures in the guarded recursive fragment:

\begin{code}
data DLast a = DLast {getDLast :: Delay (Maybe a)}

instance Stable (DLast a) where
  wait x = DLast (wait . fmap getDLast $ x)

instance Semigroup (DLast a) where
  x <> (DLast (Now Nothing)) = x
  _ <> (DLast (Now (Just y))) = DLast (Now (Just y))
  x <> (DLast (Wait y)) = 
       wait $ fmap (x  <>) (fmap DLast y)
  
instance Monoid (DLast a) where
  mempty = DLast (Now Nothing)

data DFirst a = DFirst {getDFirst :: Delay (Maybe a)}

instance Stable (DFirst a) where
  wait x = DFirst (wait . fmap getDFirst $ x)

instance Semigroup (DFirst a) where
  (DFirst (Now Nothing)) <> x = x
  (DFirst (Now (Just y))) <> _ = DFirst (Now (Just y))
  (DFirst (Wait y)) <> x = 
       wait $ fmap (<> x) (fmap DFirst y)
  
instance Monoid (DFirst a) where
  mempty = DFirst (Now Nothing)
\end{code}

These can be verified to obey the monoid laws -- identity, associativity, etc. However, they are not "infinite-associative" -- in particular, for the \hs{DFirst} monoid, the infinite chain of multiplications "$x_0 \otimes (x_1 \otimes (x_2 ...$" is $x_0$, while the re-associated chain "$... x_{\omega - 2}) \otimes x_{\omega -1}) \otimes x_\omega$" is $\bot$. The dual situation holds for the \hs{DLast} monoid. As a result, \hs{Writer DFirst} yields a non-$\bot$ writer "log" on a prompt traversal, while \hs{Writer DLast} does so on a "co-prompt" traversal.

Further, traversals also naturally arise of "mixed-promptness." For example, it is common to represent lists which can be efficiently appended to on either side as pairs of lists -- one in "forwards" order, and one in "backwards" order. Lifted to the guarded recursive fragment, we get the following, with a mixed-promptness traversal.

\begin{code}
data Bistream a = Bistream (Stream a) (Stream a) 

bicons x (Bistream xs ys) =
     Bistream (Cons x (Later xs)) ys

bisnoc y (Bistream xs ys) =
     Bistream xs (Cons y (Later ys))

instance ITraversable Bistream where
  isequence (Bistream x y) = 
       Bistream <$> isequence x <*> ibackquence y
\end{code}

This models accurately in the guarded recursive fragment the situation where traversal with writer of either or both of \hs{First} and \hs{Last} may yield a non-$\bot$ answer, depending on the manner in which the \hs{Bistream} was constructed. 

What these examples all show is that in the presence of infinite structures, standard set-theoretic equational reasoning can frequently break down. However, they also show that reasoning in the guarded-recursive fragment can explain and account for these failures, offering a fine-grained account of the varying demand-driven causal relationships between different portions of structures in a non-strict functional language.

% \section{Modal tracking of the guarded fragment}

\section{Related Work}

While the current work draws inspiration and results from many sources, we are not aware of anything else that tackles either the specific problem at hand (infinite traversals) nor the general approach of providing internal equational reasoning about partiality in a partial language. This approach is most similar to the "fast and loose" work of Danielsson, Huges, et al. \cite{danielsson2006fast}, which also developed (more formal) techniques for extracting equational principles from a total to a partial setting. Aside from the formality of the setup, the major difference is that while they extract from a standard total setting, we extract from a guarded recursive one. In a way, this is also the opposite approach to Atkey and McBride \cite{atkey2013productive} who also investigate techniques for mixing total and guarded reasoning, but instead treat the guarded recursive fragment as an "annotated-partial" embedded language which in turn allows "running" to extract back to a total host language.

Traversals were first introduced in 2008 by McBride and Paterson \cite{mcbride2008applicative} and their properties in finite settings have been explored in work by Bird, Gibbons, Oliveira, Jaskelioff, O'Connor, and Rypacek among others \cite{gibbons2009essence, jaskelioff2012investigation, bird2013understanding, jaskelioff2015representation}, with relevant results cited in this paper \emph{passim}. While Gibbons and Oliveira characterized traversals as "the essence of the iterator pattern" we believe the current work shows how infinite traversals are in a sense the essence of stream transducers. We know of no other work which investigates traversals in non-finite settings.

Use of coinductive types and a "Delay" monad to model partial computation (as we adopt in this paper) was studied by Capretta \cite{capretta2005general}. Varying quotients of this monad by weak bisimilarity have been constructed by Uustalu and Veltri \cite{uustalu2017delay}, Chapman, Uustalu, and Veltri \cite{chapman2019quotienting}, and Altenkirch, Danielsson and Kraus \cite{altenkirch2017partiality}. Bisimulation as a notion of equality for guarded recursive types has been studied by Møgelberg and Veltri \cite{mogelberg-veltri:2019} and implemented in guarded cubical Agda by Veltri and Vezzosi \cite{veltri-vezzosi:2020}. In that work a more standard construction of bisimulation is shown to coincide with path-equality on final coalgebras, which can be seen as "evaluating" a transition system, which in turn allows proofs of bisimulation to be conducted through simple equational reasoning and guarded recursion. This is similar in spirit to our approach, though we omit the construction of an equivalence to standard bisimulation entirely, as is is extraneous to the particular constructions here needed. We further noted that our bisimulation-via-evaluation approach is conducted through something resembling logical relations. A more formal relationship between bisimulation and logical relations has been explored recently in work by Hur, Dreyer et al. \cite{hur2012marriage} and Hermida, Reddy, et al. \cite{hermida_reddy_robinson_santamaria_2022}. To our knowledge, prior work has not married bisimilarity with homotopy-theory-inspired weak equivalences as in done with the gwbeq and ebeq constructions in the current paper.

Nakano's \hs{Later} modality \cite{nakano2000modality} has been the subject of much work, particularly with regards to categorical models. While most of that work is not directly relevant to the current paper, we believe it is appropriate to enumerate some highlights for those who wish to explore further. The modality's connection to Gödel-Löb provability logic and Kripke models of such was explored further by in Nakano in the next year, \cite{nakano2001fixed}, and more recently the connection to intensional reasoning has been pursued in work by Kavvos \cite{kavvos2017intensionality, kavvos2017semantics} and Chen and Ko \cite{chen-ko:2022}.

A significant extension of the modality to "clocked" type theories with multiple clocks and "running" of computations of finite depth (for the purposes of modeling coinductive types) was given by Atkey and McBride \cite{atkey2013productive}. Other related extensions have been proposed by Birkedal, Grathwohl et al. \cite{birkedal2017guarded} and Guatto \cite{guatto:2018}. While clocks allow running of \emph{finite} computations to extract total data, the \hs{leval} construction in this paper instead is built in a non-total setting, and allows running of all computations, with results taking domain semantics.

A class of categorical models ("synthetic guarded domain theory") of the modality in the topos of trees, and more generally sheaves over complete Heyting algebras was introduced in 2011 by Birkedal, Møgelberg et al. \cite{birkedal2011first}. Since then, much subsequent work has been done on extending this approach to models to handle more general type theories, including clocked theories, dependent theories, and cubical theories. Such models are an active area of current research, and notable contributions have been made by (among others) Birkedal, Bizjak, Grathwohl, Gratzer, Harper, Møgelberg, Palombi, and Sterling. \cite{gratzer-birkedal:2022, sterling-harper:2018, bbcgsv:2019, bizjak2020denotational, palombi-sterling:2022}.

Use of synthetic guarded domain theory as a tool to formalize denotational semantics has been particularly explored in work by Paviotti \cite{paviotti:2016, mogelberg-paviotti:2016, paviotti-mogelberg-birkedal:2015}.

Unlike the above-discussed work pertaining to guarded recursion, our work emphasizes equational reasoning in a productive fragment of a language with general recursion, and extraction of principles and results to the full language.

\section{Future Work and Conclusion}

The current paper has built a sublanguage internal to Haskell which allows for equational reasoning in a semantics richer than $\mathbf{Set}$ -- and in particular, in the presence of potentially nonterminating computation. In this setting, facts can be derived about the productivity of various Haskell functions, and extracted back to the full language. This has allowed a simple characterization of when an infinite traversal is productive, as well as fine-grained intensional analysis of the tricky phenomena of specific infinite traversals. But many questions remain.

The toolkit developed is optimized for lightweight, semi-formal reasoning. However, it could stand to be formalized more rigorously, which would require its construction in an established domain-theoretic setting -- either guarded, or otherwise.  Further, the analysis carried out here of infinite traversals with \hs{State} remains somewhat unsatisfactory. Perhaps it is an example where a more fine-grained system such as the "time-warps" in \cite{guatto:2018} is necessary for capturing fully the desired semantics.

Many other type class laws and characterizations in Haskell besides \hs{Traversable} are also mainly defined and studied on the total fragment of the language. It is worth exploring if the approach of this paper can extend to analyzing other type class operations in the case of partial or infinite computation as well.

The typeclass-encoded bisimulation-via-evaluation construction in this paper can be seen as internalizing the method of logical relations, with induction on types pairing terms with witnesses to their semantics. It would be interesting to generalize and solidify this technique, relating it to existing literature.

An earlier approach to the questions here made use of the interleaved effect algebras of \cite{interleaving, induction-with-effects}. While that yielded unsatisfactory results, perhaps those algebras could be coupled in a more principled way with "causality-as-an-effect" by explicitly considering them in relation to guarded recursion.

A further question to ask is if there is a single structure that can play the role with infinite traversals that lists do with finite traversals. In particular, as described in Bird, Gibbons et al. \cite{bird2013understanding} finite traversable functors can be decomposed into a "structure" signature, and a "contents" list. If one swaps out "list" for something else (say, a stream, or a stream of lists), is the same possible for infinite traversals? It seems likely that the "free guarded magma" (i.e. the \hs{ITree} functor considered early on) would play such a role.

Then there are more general questions regarding category theoretic models. The current work has no equivalent of the classical result relating traversable functors to finitary polynomial functors. It is not clear precisely what one might conjecture the analogue of this would be in a guarded recursive setting. Many of the equivalent characterizations of polynomial functors only hold for endofunctors over $\mathbf{Set}$ \cite{spivak2022reference}. A first step would be determining what can still be said of these characterizations in the topos of trees. It would also be necessary to characterize those guarded recursive datatypes that arise as liftings of recursive datatypes from a total setting -- it seems likely that the condition would shift from overall finiteness to "pointwise" (or "stepwise") finiteness.

As far as general categorical speculation -- the guarded fragment itself appears to be a comonadic modality in the partial language, and could perhaps be captured as a comonadic coreflective subcategory. This raises the connection of it bears any special relationship with the \hs{Delay} monad -- and more generally if a factorization system arises. A tool in this analysis could be the gwbeq construction, which gives a system of weak equivalences. Further to this, it would be interesting to explore what arises if one quotients a guarded-recursive model by gwbeq, taking the "homotopy category" -- this could be a path to useful models of denotational semantics.

\begin{acks}
This work grew out of some early discussions with Edward Kmett and Dan Doel that posed the questions that motivate it. The basic idea took shape after a useful discussion with Bob Atkey, after which it simmered for some years. Special thanks are due to James Deikun for many useful ideas, not least suggesting the name "Predictable". Callan McGill and Jeff Polakow both provided comments on the draft. Finally, we thank the Haskell Symposium reviewers for detailed and helpful comments and suggestions. 
\end{acks}

% TODO -- anything to say about geometric logic or fast exhaustive searchable sets?

\newpage
 \nocite{*}
 \bibliography{predictable}

\begin{comment}
\begin{code}    


z :: [Writer (DLast Int) Int]  
z = [tellit 1, tellit 2, tellit 3]
  where tellit x = tell (DLast . Now $ (Just x)) >> pure x
  
zx = Bistream (listToStream  (cycle z)) (listToStream (cycle $ drop 1 z))  

zz :: [Writer (DFirst Int) Int]  
zz = [tellit 17, tellit 1, tellit 2, tellit 3]
  where tellit x = tell (DFirst . Now $ (Just x)) >> pure x
  
zzx = Bistream (listToStream  (cycle zz)) (listToStream (cycle $ drop 1 zz))  

zzz :: [Writer (PStream Int) Int]
zzz = [tellit 1, tellit 2, tellit 3]
  where tellit x = tell (PCons x PNil) >> pure x
  
quux :: [State Int Int]
quux = [get]



deriving instance Show a => Show (Delay a)
deriving instance Functor Stream
  

\end{code}
\end{comment}

\end{document}